## Entanglement dynamics of a two-qutrits system coupled to a spin chain

S. M. Moosavi Khansari[1] . F. Kazemi Hasanvand[1]

[1] Department of Physics, Faculty of Basic Sciences, Ayatollah Boroujerdi University, Boroujerd, Iran;

Corresponding Author E-mail: m.moosavikhansari@abru.ac.ir
E-mail: fa_kazemi270@yahoo.com

**Abstract.** In this paper, we investigate the entanglement dynamics of a two qutrits system interacting with a spin environment. Using negativity as the entanglement measure, we study the entanglement dynamics of the system. The calculations show that in cases where the entanglement decays quickly, the environment will have a quantum phase transition.

## 1 Introduction

Entanglement is a fundamental and key aspect of quantum mechanics that plays an essential role in the realms of information processing and quantum computing. In recent years, this intriguing phenomenon has been the subject of extensive study and investigation by many researchers in the field [1]. Given that quantum systems inevitably interact with their surroundings, it becomes increasingly important to thoroughly understand the ways in which the environment influences entanglement. Additionally, developing effective methods to control this entanglement during interactions between systems and their environments is of utmost significance [2,3]. Recent research efforts have notably focused on examining the environmental effects specifically on qubit-qubit, qutrit-qutrit, and qubit-qutrit spin systems [4-11]. In this article, we will first introduce and present the complete Hamiltonian for a quantum system that consists of two qutrits. These qutrits are under the influence of a surrounding spin chain that demonstrates threefold interactions. Following this introduction, we will apply the negativity criterion in order to conduct a detailed analysis of the dynamics associated with entanglement [12-14]. Furthermore, we will investigate how these dynamics depend on various parameters, thereby enhancing our understanding of the complex interplay between entanglement and environmental factors.

## 2 Theoretical calculations

The system comprises two non-interacting qutrits, each influenced by a spin chain with three-way interactions. The Hamiltonian of the system is expressed as follows:

$$\mathrm{H_E} = -\sum_{\ell=1}^{n}\left(\frac{1+\gamma}{2}\sigma_{\ell}^{x}\sigma_{\ell+1}^{x} + \frac{1-\gamma}{2}\sigma_{\ell}^{y}\sigma_{\ell+1}^{y}\right) - \eta\sum_{\ell=1}^{n}\sigma_{\ell}^{z} - \qquad (1)$$

$$\sum_{\ell=1}^{n}\alpha(\sigma_{\ell+1}^{x}\sigma_{\ell}^{z}\sigma_{\ell-1}^{y}+\sigma_{\ell+1}^{y}\sigma_{\ell}^{z}\sigma_{\ell-1}^{x})$$

$$H_I=-(g_A S_A^z+g_B S_B^z)\sum_{\ell=1}^{n}\sigma_{\ell}^{z} \qquad (2)$$

Here, n represents the number of particles in the spin chain, while $S_A^z$ and $S_B^z$ are the z-direction spin operators for each qutrit. $g_A$ and $g_B$ denote the coupling strengths of spins A and B with the environment, $\alpha$ is the triplet interaction strength, $\eta$ represents the magnetic field intensity, and $\gamma$ signifies the anisotropy. We assume the initial state of the system is as follows:

$$|\psi\rangle_s=\frac{1}{\sqrt{3}}(|00\rangle+|11\rangle+|22\rangle) \qquad (3)$$

Thus, the density matrix for the system's initial state is expressed as follows:

$$\rho_s(0)=\begin{pmatrix}\frac{1}{3}&0&0&0&\frac{1}{3}&0&0&0&\frac{1}{3}\\0&0&0&0&0&0&0&0&0\\0&0&0&0&0&0&0&0&0\\0&0&0&0&0&0&0&0&0\\\frac{1}{3}&0&0&0&\frac{1}{3}&0&0&0&\frac{1}{3}\\0&0&0&0&0&0&0&0&0\\0&0&0&0&0&0&0&0&0\\0&0&0&0&0&0&0&0&0\\\frac{1}{3}&0&0&0&\frac{1}{3}&0&0&0&\frac{1}{3}\end{pmatrix} \qquad (4)$$

The transformed density matrix of the system, resulting from its interaction with the environment, is expressed as follows:

$$\rho_s(t)=\frac{1}{3}(|00\rangle\langle 00|+|11\rangle\langle 11|+|22\rangle\langle 22|+F_{15}|00\rangle\langle 11|+F_{19}|00\rangle\langle 22|+F_{15}{}^{*}|11\rangle\langle 00|+ \qquad (5)$$

$$F_{59}|11\rangle\langle 22|+F_{19}{}^{*}|22\rangle\langle 00|+F_{59}{}^{*}|22\rangle\langle 11|)$$

where the decoherence coefficients, are obtained from the following relationship:

$$F_{\mu\nu}=$$

$$\prod_{k=1}^{M}e^{it(\xi_k^{\lambda_\mu}-\xi_k^{\lambda_\nu})}((1-e^{2it\xi_k^{\lambda_\mu}})(1-e^{2it\xi_k^{\lambda_\nu}})\sin(\frac{1}{2}(\theta_k^{\lambda_\mu}-\theta_k^{\eta}))\sin(\frac{1}{2}(\theta_k^{\lambda_\nu}-\theta_k^{\eta}))\cos(\frac{1}{2}(\theta_k^{\lambda_\mu}- \qquad (6)$$

$$\theta_k^{\lambda_\nu}))-((1-e^{2it\xi_k^{\lambda_\mu}})\sin^2(\frac{1}{2}(\theta_k^{\lambda_\mu}-\theta_k^{\eta})))-(1-e^{2it\xi_k^{\lambda_\nu}})\times\sin^2(\frac{1}{2}(\theta_k^{\lambda_\nu}-\theta_k^{\eta}))+1)$$

and also the magnitude of decoherence coefficients, $|F_{\mu\nu}|$, are obtained from the following relationship:

$$|F_{\mu\nu}| = \prod_{k=1}^{M}(-4\sin(\theta_k^{\eta}-\theta_k^{\lambda_\mu})\sin(\theta_k^{\eta}-\theta_k^{\lambda_\nu})\sin^2(\frac{1}{2}(\theta_k^{\lambda_\mu}-\theta_k^{\lambda_\nu}))\sin^2(t\,\Lambda_k^{\lambda_\mu})\sin^2(t\,\Lambda_k^{\lambda_\nu})+2\sin(\theta_k^{\eta}-\theta_k^{\lambda_\mu})\sin(\theta_k^{\eta}-\theta_k^{\lambda_\nu})\sin(t\,\Lambda_k^{\lambda_\mu})\sin(t\,\Lambda_k^{\lambda_\nu})\cos(t\,\Lambda_k^{\lambda_\mu}-t\,\Lambda_k^{\lambda_\nu})-\sin^2(\theta_k^{\eta}-\theta_k^{\lambda_\mu})\sin^2(t\,\Lambda_k^{\lambda_\mu})-\sin^2(\theta_k^{\eta}-\theta_k^{\lambda_\nu})\sin^2(t\,\Lambda_k^{\lambda_\nu})+1)^{1/2} \quad (7)$$

in these relationships we use

$$\xi_k^{\lambda_\mu} = 2\sqrt{\gamma^2\sin^2(\frac{2\pi k}{n})+(\lambda_\mu-\cos(\frac{2\pi k}{n}))^2)} \quad (8)$$

and

$$\Lambda_k^{\lambda_\mu} = 2(\alpha\sin(\frac{4\pi k}{n})+\sqrt{\gamma^2\sin^2(\frac{2\pi k}{n})+(\lambda_\mu-\cos(\frac{2\pi k}{n}))^2)} \quad (9)$$

and $\lambda_\mu$ for $\mu = 1,2,\ldots,9$ are defined as follows

$$\lambda_1 = \eta+g_A+g_B,\ \lambda_2 = \eta+g_A,\ \lambda_3 = \eta+g_A-g_B \quad (10)$$

$$\lambda_4 = \eta+g_B, \quad \lambda_5 = \eta, \quad \lambda_6 = \eta-g_B$$

$$\lambda_7 = \eta-g_A+g_B, \quad \lambda_8 = \eta-g_A, \quad \lambda_9 = \eta-g_A-g_B$$

The following relationships can be defined

$$\theta_k^{\lambda_\mu} = \tan^{-1}\left(\frac{\gamma\sin(\frac{2\pi k}{n})}{\lambda_\mu-\cos(\frac{2\pi k}{n})}\right) \quad (11)$$

and

$$\theta_k^{\eta} = \tan^{-1}\left(\frac{\gamma\sin(\frac{2\pi k}{n})}{\eta-\cos(\frac{2\pi k}{n})}\right) \qquad (12)$$

in these relationships, $M = (n - 1) / 2$. Based on these relations, it can be expressed as

$$\rho_s(t) = \begin{pmatrix} \frac{1}{3} & 0 & 0 & 0 & \frac{F_{15}}{3} & 0 & 0 & 0 & \frac{F_{19}}{3} \\ 0 & 0 & 0 & 0 & 0 & 0 & 0 & 0 & 0 \\ 0 & 0 & 0 & 0 & 0 & 0 & 0 & 0 & 0 \\ 0 & 0 & 0 & 0 & 0 & 0 & 0 & 0 & 0 \\ \frac{F_{15}^*}{3} & 0 & 0 & 0 & \frac{1}{3} & 0 & 0 & 0 & \frac{F_{59}}{3} \\ 0 & 0 & 0 & 0 & 0 & 0 & 0 & 0 & 0 \\ 0 & 0 & 0 & 0 & 0 & 0 & 0 & 0 & 0 \\ 0 & 0 & 0 & 0 & 0 & 0 & 0 & 0 & 0 \\ \frac{F_{19}^*}{3} & 0 & 0 & 0 & \frac{F_{59}^*}{3} & 0 & 0 & 0 & \frac{1}{3} \end{pmatrix} \qquad (13)$$

We use the negativity criterion to calculate entanglement. For a quantum state with density matrix $\rho$, the negativity is defined as follows:

$$N(\rho) = \frac{1}{2}\left(\left\|\rho^{T_i}\right\|_1 - 1\right) \qquad (14)$$

where $\rho^{T_i}$ is the partial transpose of $\rho$ with respect to i and $\|\cdot\|_1$ denotes the trace norm.

Thus, for the state outlined in relation (3), if we designate i as subsystem A, it can be expressed as follows

$$\rho^{T_A} = \begin{pmatrix} \frac{1}{3} & 0 & 0 & 0 & 0 & 0 & 0 & 0 & 0 \\ 0 & 0 & 0 & \frac{F_{15}^*}{3} & 0 & 0 & 0 & 0 & 0 \\ 0 & 0 & 0 & 0 & 0 & 0 & \frac{F_{19}^*}{3} & 0 & 0 \\ 0 & \frac{F_{15}}{3} & 0 & 0 & 0 & 0 & 0 & 0 & 0 \\ 0 & 0 & 0 & 0 & \frac{1}{3} & 0 & 0 & 0 & 0 \\ 0 & 0 & 0 & 0 & 0 & 0 & 0 & \frac{F_{59}^*}{3} & 0 \\ 0 & 0 & \frac{F_{19}}{3} & 0 & 0 & 0 & 0 & 0 & 0 \\ 0 & 0 & 0 & 0 & 0 & \frac{F_{59}}{3} & 0 & 0 & 0 \\ 0 & 0 & 0 & 0 & 0 & 0 & 0 & 0 & \frac{1}{3} \end{pmatrix} \qquad (15)$$

We can calculate the trace norm of this matrix by first determining the absolute values of its eigenvalues

$$\left\{\frac{1}{3},\frac{1}{3},\frac{1}{3},\frac{2}{3}|F_{15}|,\frac{2}{3}|F_{19}|,\frac{2}{3}|F_{59}|\right\} \qquad (16)$$

Finally, the negativity is derived as follows:

$$N=\frac{(\frac{1}{3}+\frac{1}{3}+\frac{1}{3}+\frac{2}{3}|F_{15}|+\frac{2}{3}|F_{19}|+\frac{2}{3}|F_{59}|)-1}{2} \qquad (17)$$

After simplification, we obtain the following relationship:

$$N=\frac{\left(1+\frac{2}{3}|F_{15}|+\frac{2}{3}|F_{19}|+\frac{2}{3}|F_{59}|\right)-1}{2}=\frac{(|F_{15}|+|F_{19}|+|F_{59}|)}{3} \qquad (18)$$

## 3 Results and discussion

This article delves into the intricate dynamics of entanglement within a two-qutrits system that is influenced by an external spin environment. For the first time, we incorporate a spin chain characterized by three-way interactions into the total Hamiltonian of the system, allowing for a comprehensive analysis of the entanglement behavior. By utilizing negativity as a quantitative measure of entanglement, we have conducted a thorough examination of the dynamics that govern the entanglement within this system.

Figures 1 through 5 illustrate that entanglement experiences a rapid decrease when the parameter $\eta$ is set to 1. In contrast, Figure 6 reveals that this decrease in entanglement occurs at a significantly slower rate. Throughout Figures 1 to 6, we observe that entanglement is maximized and remains nearly constant when $\eta$ is equal to 1.2. However, when we consider the cases where $\eta$ is set to 0 and $\eta$ is set to 0.5, we find that the entanglement fluctuates around a similar value, indicating a different behavior under these conditions.

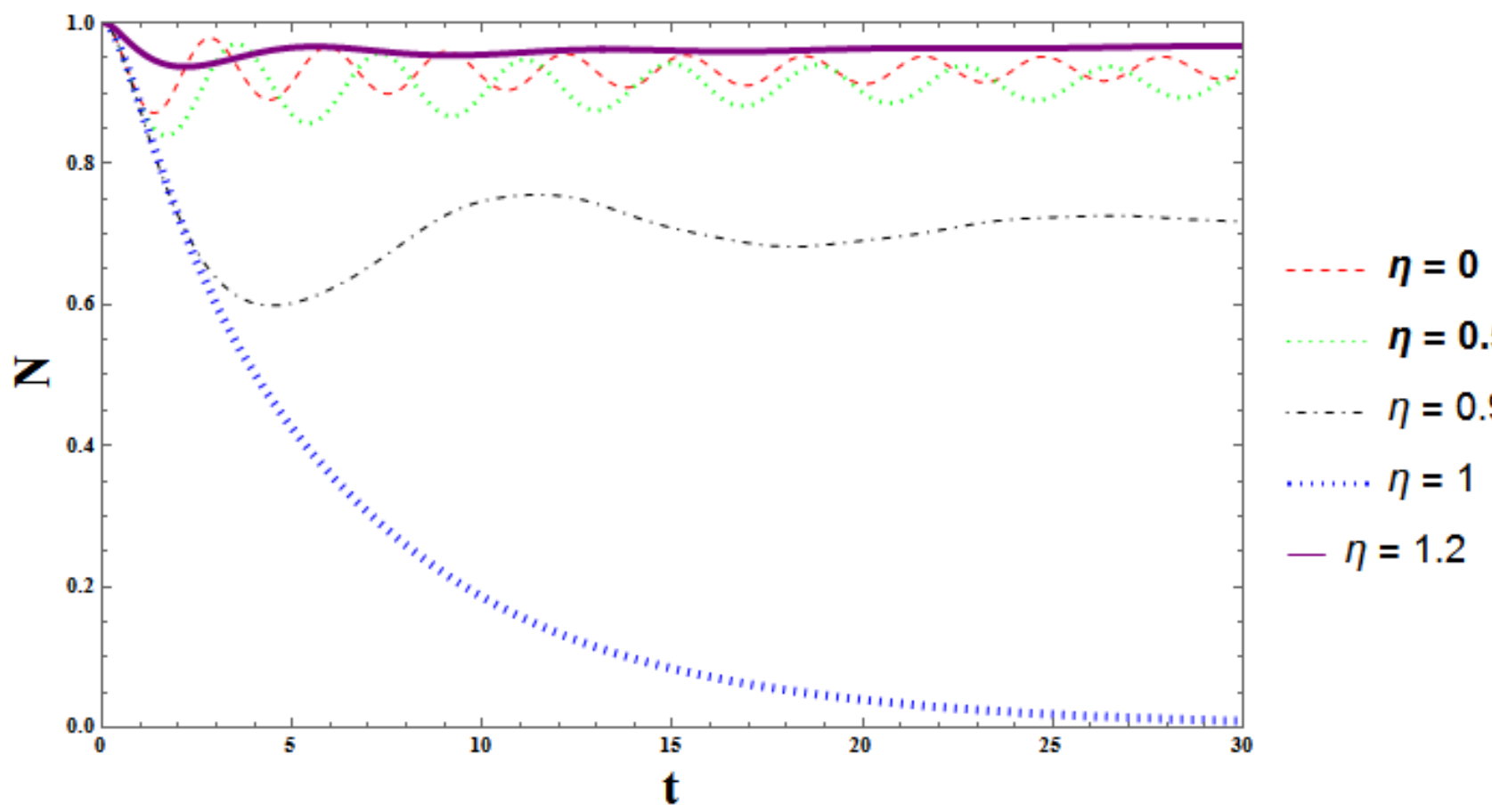

Figure 1: The negativity diagram over time for fixed parameters: $g_A = 0.005, g_B = 0.005, \gamma = 0.5, \alpha = 0.5, n = 3001$, and five different values of $\eta$.

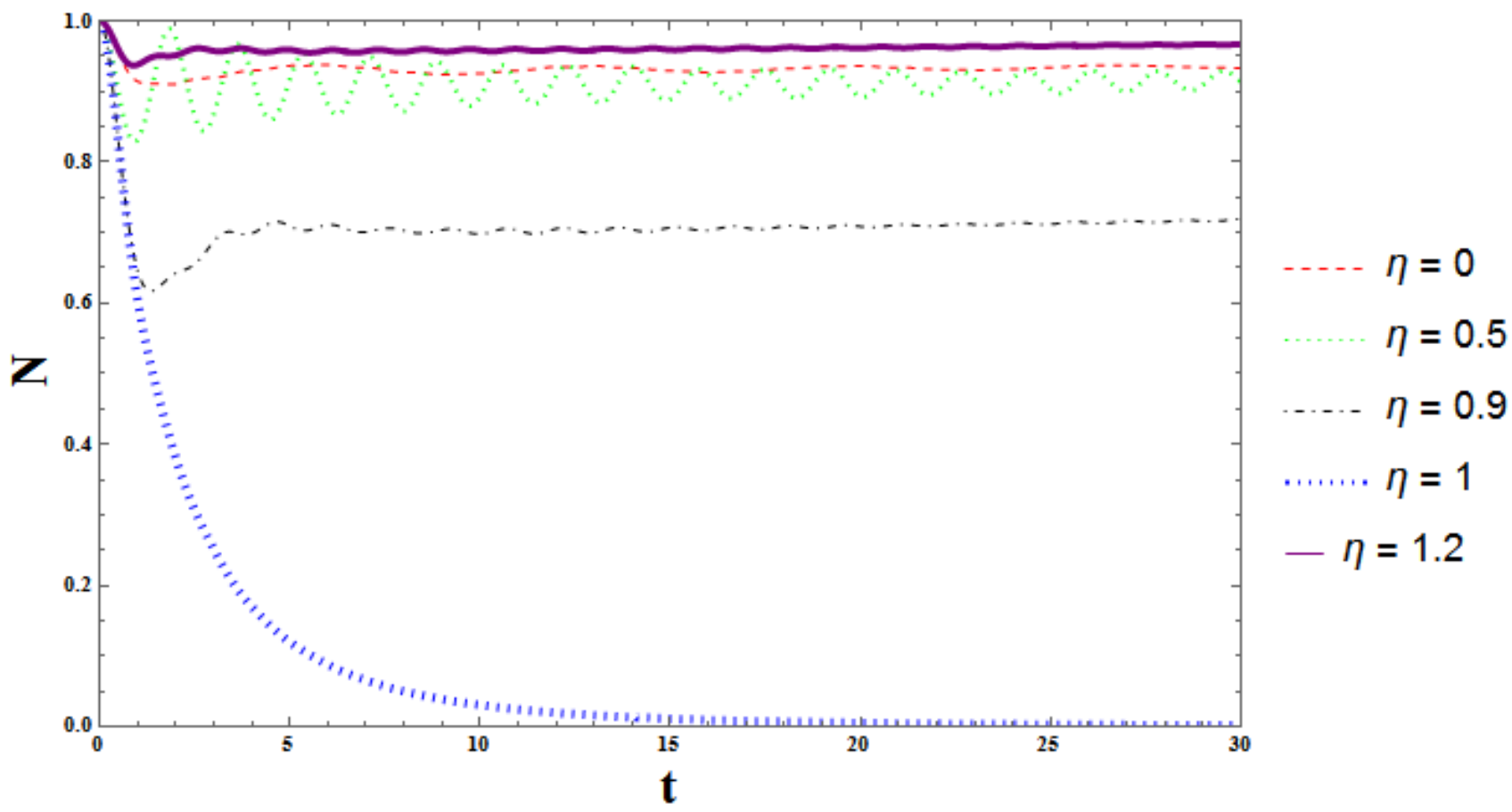

Figure 2: Negativity diagram over time for fixed parameters: $g_A = 0.005, g_B = 0.005, \gamma = 0.5, \alpha = 0.5, n = 3001$ and for five different values of $\eta$.

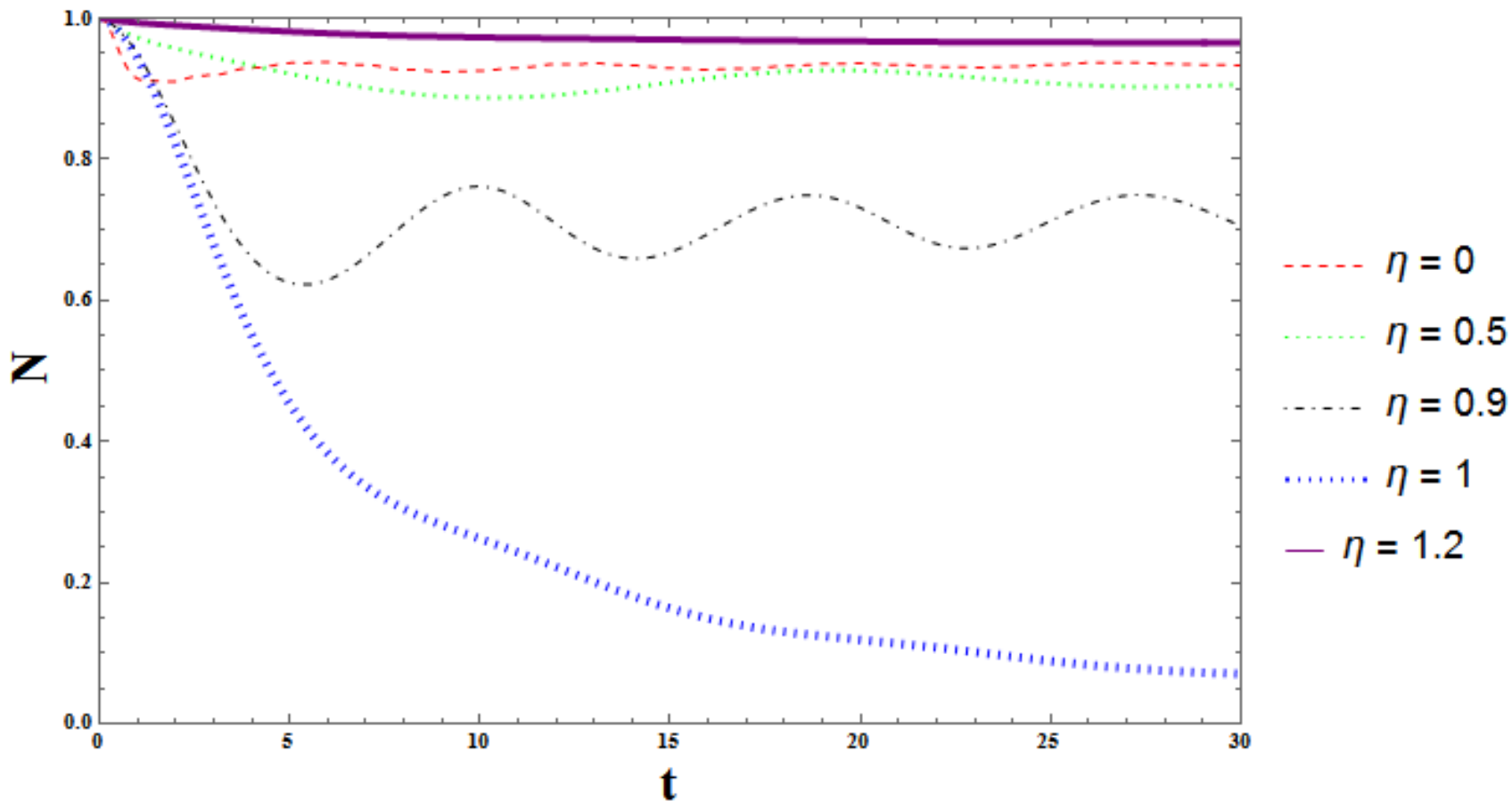

Figure 3: Graph of negativity in terms of time for fixed data $g_A = 0.005, g_B = 0.005, \gamma = 0.5, \alpha = -0.5, n = 3001$ and for five different values of $\eta$.

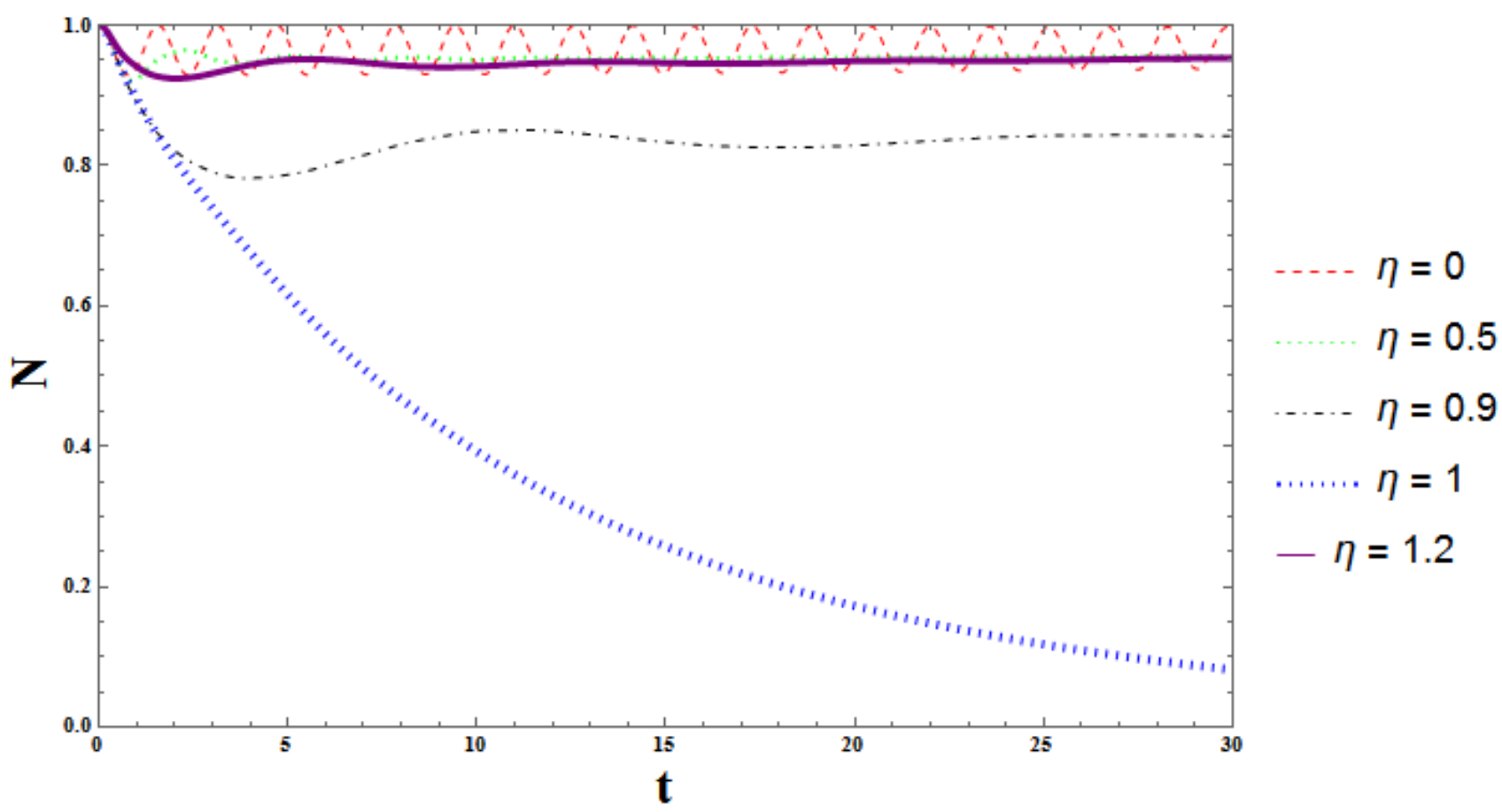

Figure 4: Negativity diagram in terms of time for fixed data $g_A = 0.005, g_B = 0.005, \gamma = 1, \alpha = 0, n = 3001$ and for five different values of $\eta$.

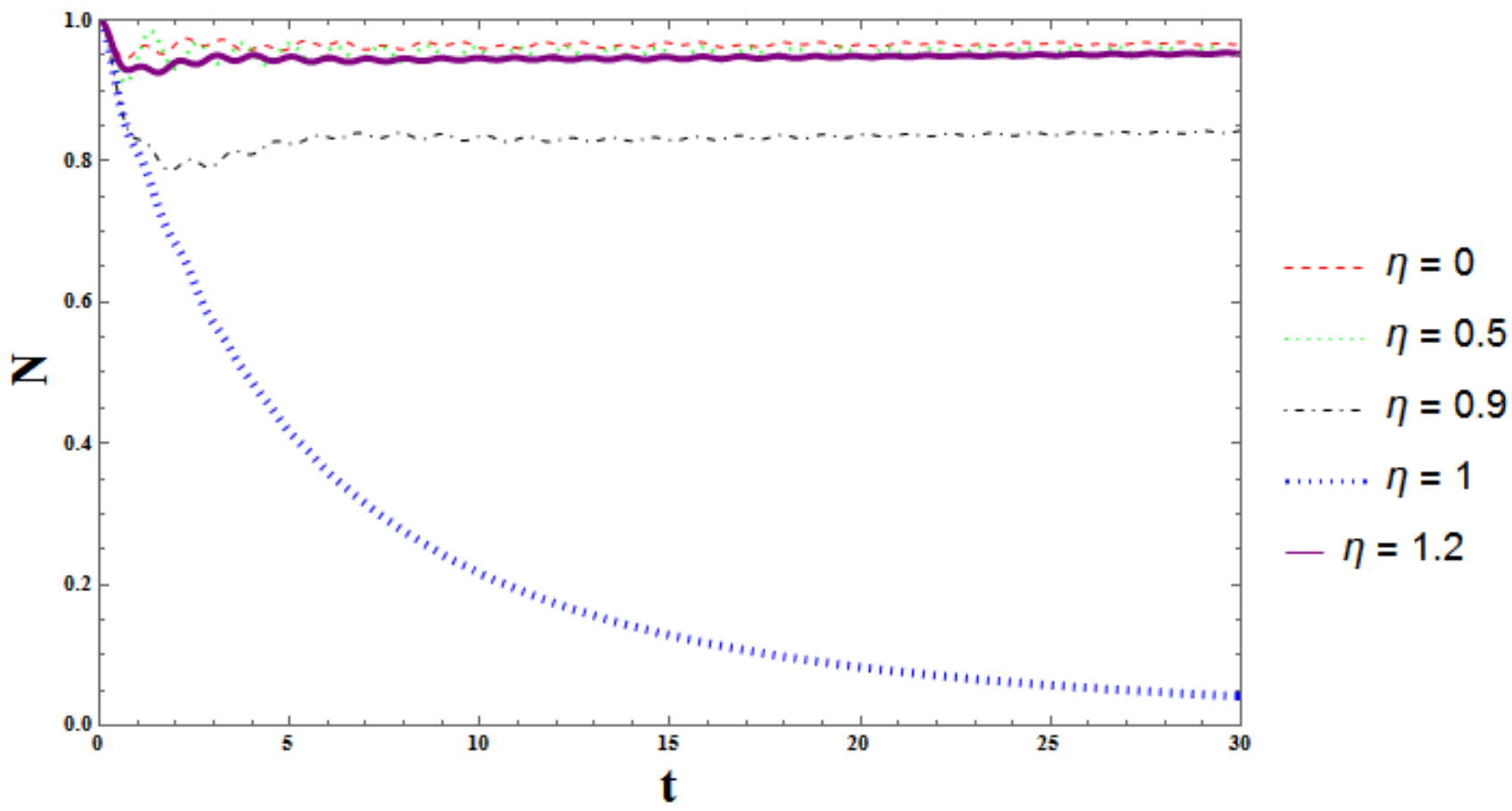

Figure 5: Negativity diagram in terms of time for fixed data $g_A = 0.005, g_B = 0.005, \gamma = 1, \alpha = 0.5, n = 3001$ and for five different values of $\eta$.

Additionally, for the case where $\eta$ is equal to 0.9, we note that entanglement decreases and shows fluctuations in certain instances. The variable $\alpha$ plays a crucial role in influencing both the damping time and the frequency of these entanglement fluctuations. The observation of rapid entanglement decay strongly suggests the presence of a quantum phase transition occurring within the environment. It is particularly noteworthy that when the parameter $\alpha$ is set to -0.5 and $\gamma$ is equal to 1, the reduction of entanglement at $\eta = 1$ is significantly slower compared to the reductions observed in other scenarios.

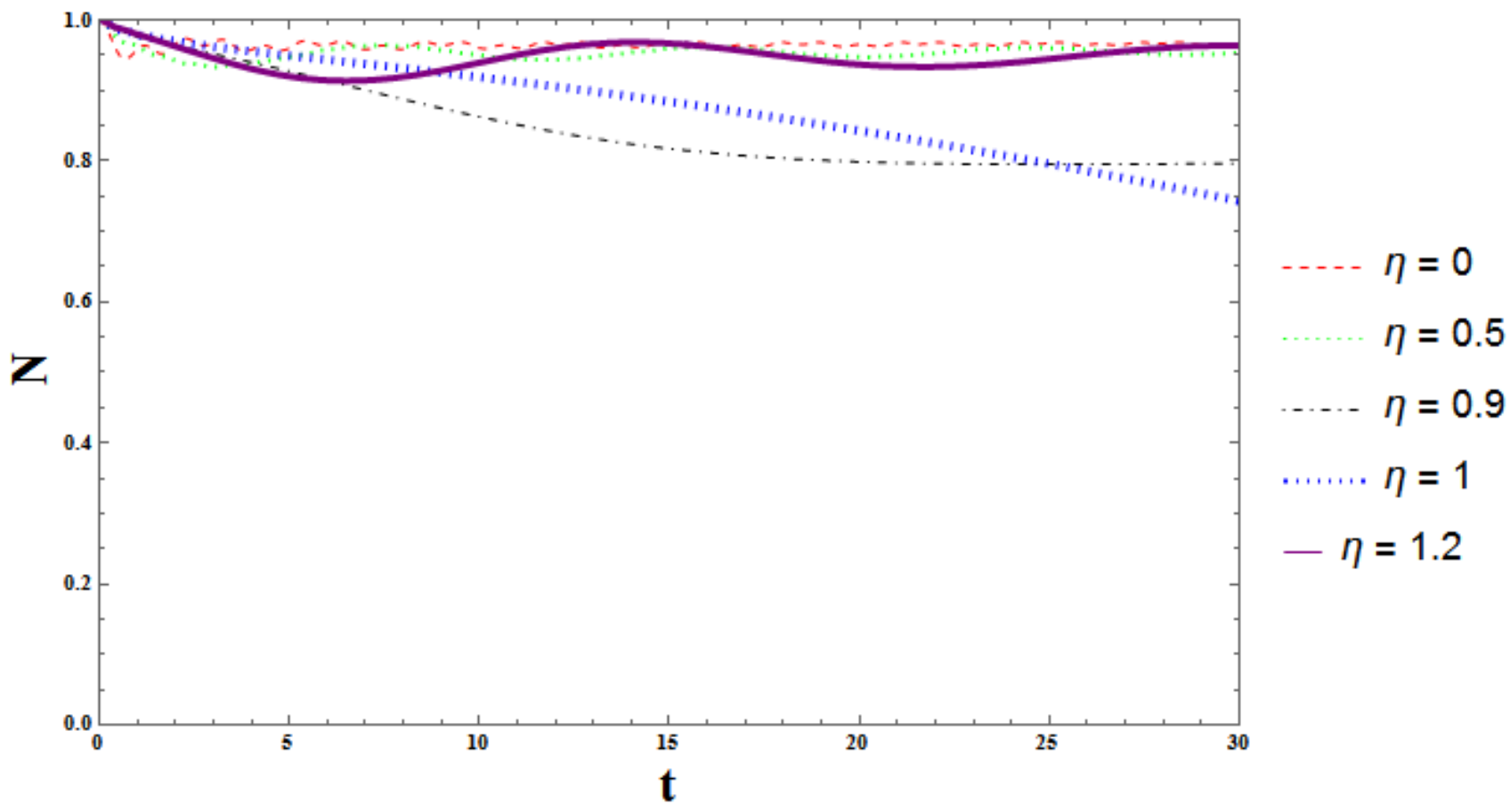

Figure 6: Negativity diagram in terms of time for fixed data $g_A = 0.005, g_B = 0.005, \gamma = 1, \alpha = -0.5, n = 3001$ and for five different values of $\eta$.

In Figure 7, we illustrate the negativity as a function of time (t) alongside the variable $\alpha$ for the case where $\gamma$ is equal to 1, which corresponds to the Ising model, at the critical magnetic field strength set at $\eta = 1$. The maximum negativity is observed at $\alpha = -0.5216$. Within the range of $-0.5216 \leq \alpha \leq 0.5$, we see that entanglement decreases rapidly. Conversely, in the range of $-1 \leq \alpha \leq -0.5216$, the time taken for this reduction increases, indicating significant delays in the decay of entanglement, particularly at the point where $\alpha$ is equal to -0.5216.

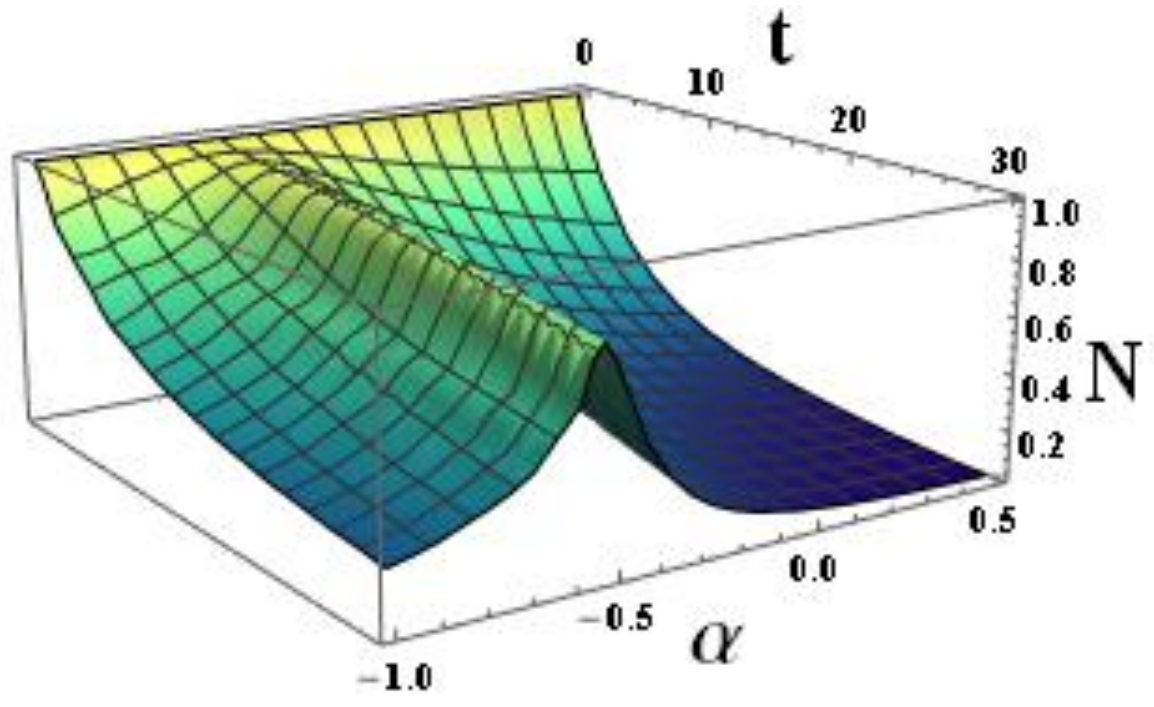

Figure 7: Negativity diagram in terms of $\alpha$ and in terms of t for $g_A = 0.005, g_B = 0.005, \gamma = 1, n = 3001, \eta = 1$ (Ising model)

Figure 8 presents negativity as a function of time and variable $\alpha$ for the case where $\gamma$ is equal to 0.5, corresponding to the XY model, also at $\eta = 1$. Here, we observe that maximum negativity occurs at $\alpha = -0.2695$. As we move away from this point, specifically within the range of $-0.2695 \leq \alpha \leq 0.5$, we witness a rapid decay in entanglement. However, when we consider the range of $-1 \leq \alpha \leq -0.2695$, the rate of decay slows down significantly when compared to the previous range. Remarkably, at $\alpha = -0.2695$, the entanglement remains stable over time, indicating a unique characteristic of this parameter setting.

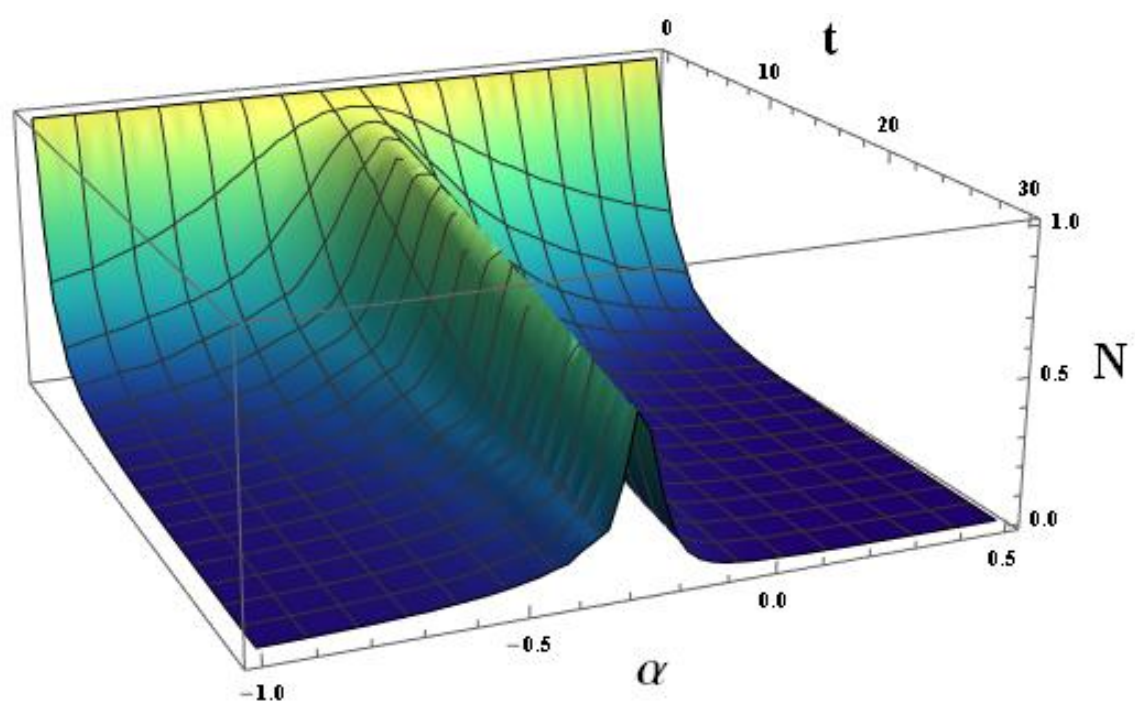

Figure 8: Negativity diagram in terms of $\alpha$ and in terms of t for $g_A = 0.005, g_B = 0.005, \gamma = 0.5, n = 3001, \eta = 1$ (XY model)

Finally, Figure 9 illustrates negativity as a function of time and variable $\alpha$ for the case where $\gamma$ is equal to 0.2, again at $\eta = 1$. The maximum negativity in this scenario is achieved at $\alpha = -0.1206$. In this particular case, we observe that the limit of $\alpha$ decreases, which subsequently leads to a delay in the reduction of entanglement. This results in a significant postponement of the decay of entanglement at the specific point where $\alpha$ is equal to -0.1206, further emphasizing the complex interplay of parameters in this system.

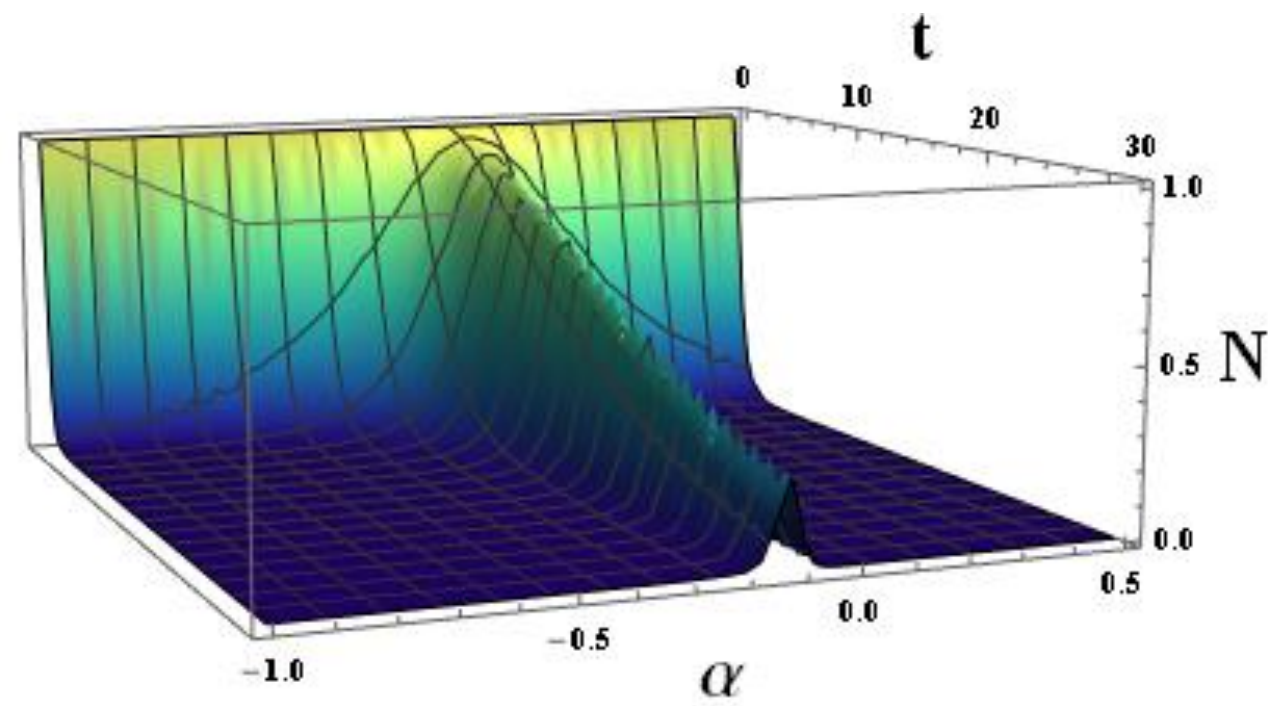

Figure 9: Negativity diagram in terms of $\alpha$ and in terms of t for $g_A = 0.005, g_B = 0.005, \gamma = 0.2, n = 3001, \eta = 1$

## 4 Hholographic negativity in two-qutrit systems

AdS/CFT, which stands for the Anti-de Sitter/Conformal Field Theory correspondence, is a profound theoretical framework that was introduced by the physicist Juan Maldacena in the year 1997. This correspondence elucidates a fascinating relationship between two distinct yet interconnected theories: Anti-de Sitter Space (AdS) and Conformal Field Theory (CFT). Anti-de Sitter Space is characterized as a universe model that possesses a constant negative curvature, making it particularly relevant in the realms of string theory and quantum gravity. On the other hand, Conformal Field Theory is a type of quantum field theory that remains invariant under conformal transformations, which are transformations that preserve angles but not necessarily distances. The essence of the correspondence posits that a gravitational theory formulated within the context of AdS space has a direct correspondence to a CFT that exists on the boundary of this space. This duality not only enhances our understanding of complex phenomena such as quantum gravity and black holes but also sheds light on the intricate nature of quantum entanglement. As such, it has emerged as a vital and foundational concept in the landscape of modern theoretical physics, influencing various areas of research and inquiry. Entanglement entropy serves as a crucial measure of quantum entanglement present within a given system, providing insights into how much information is effectively lost when a quantum system is divided into two distinct parts. In the framework of quantum mechanics, entanglement arises when particles become correlated in such a way that their individual states become interdependent. This interdependence is quantified through the concept of entanglement entropy, which serves to encapsulate the degree of correlation between the subsystems. For a bipartite system that consists of two subsystems, labeled A and B, the entanglement entropy S can be calculated using the reduced density matrix $\rho_A$ associated with subsystem A. The formula for this calculation is expressed as follows: $\mathrm{S(A)} = -\mathrm{Tr}(\rho_\mathrm{A} \log \rho_\mathrm{A})$, where Tr denotes the trace operation and log represents the logarithm function. A higher value of entanglement entropy indicates a stronger correlation between the subsystems, while a value of zero entropy signifies the absence of any entanglement. The concept of entanglement entropy is of paramount importance in various fields, including quantum information theory, condensed matter physics, and quantum gravity. It plays a particularly significant role in the study of black holes, especially in relation to the information paradox and the holographic principle. Understanding entanglement entropy is essential for gaining deeper insights into the behavior of quantum systems and the intricate correlations that exist within them. Holographic negativity is another important measure of entanglement in quantum systems, and it proves to be particularly useful for comprehending how information is encoded within holographic theories. In the context of a two-qutrit system, holographic negativity serves to quantify the entanglement that exists between two subsystems, each of which can occupy one of three possible states. The ongoing research focused on holographic negativity in two-qutrit systems aims to deepen our understanding of the intricate connection that exists between entanglement and the geometric properties inherent in holographic codes. These studies meticulously analyze how the entanglement present between the two qutrits is intricately related to the geometric characteristics of the corresponding holographic code. The overarching goal of this research is to explore the dynamics and behaviors that emerge from the interplay between entanglement and geometry within these systems. This exploration involves the calculation of logarithmic negativity, which is recognized as a key measure of entanglement, particularly in the context of quantum error-correcting codes. Through this research, scientists aim to uncover new insights into the fundamental nature of quantum entanglement and its relationship with geometric structures, thereby contributing to the broader understanding of quantum information and its implications in various theoretical frameworks. In this paper, we consider a two-qutrits system coupled to a spin chain and study the entanglement dynamics via negativity. Our calculation is a direct approach and can be compared with the holographic method result to confirm it. Detailed explanations can be found in references [15-17]. Holographic negativity is used to study entanglement in quantum systems, especially in the context of holographic theories. Its applications include:

**Understanding Quantum Information**: It helps researchers understand how information is stored and processed in quantum systems.

**Analyzing Quantum States**: Holographic negativity can measure the entanglement between different parts of a quantum system, providing insights into their relationships.

**Studying Quantum Error Correction**: It plays a role in developing quantum error-correcting codes, which are essential for maintaining the integrity of quantum information.

**Exploring Black Holes**: Researchers use holographic negativity to investigate the properties of black holes and their connection to quantum mechanics.

**Linking Geometry and Entanglement**: It helps explore how the shape and structure of space relate to the entanglement in quantum systems, enhancing our understanding of both fields.

## 5 Conclusion

This article explores the entanglement dynamics of a two-qutrits system within a spin environment, integrating a spin chain with three-way interactions into the Hamiltonian. We use negativity as a measure of entanglement. Figures 1 to 5 reveal a rapid decrease in entanglement for $\eta = 1$, while Figure 6 shows a slower decline. At $\eta = 1.2$, entanglement is maximized, but it fluctuates at $\eta = 0$ and $\eta = 0.5$. For $\eta = 0.9$, entanglement decreases with fluctuations. The variable $\alpha$ influences the damping time and fluctuation frequency, with rapid decay indicating a quantum phase transition. Notably, for $\alpha = -0.5$ and $\gamma = 1$, the reduction of entanglement at $\eta = 1$ is slower. Figure 7 depicts negativity over time and variable $\alpha$ for $\gamma = 1$ (Ising model) at $\eta = 1$, peaking at $\alpha = -0.5216$. For $-0.5216 \leq \alpha \leq 0.5$, entanglement decreases rapidly, while $-1 \leq \alpha \leq -0.5216$ shows longer reduction times and delays at $\alpha = -0.5216$. Figure 8 presents negativity for $\gamma = 0.5$ (XY model) at $\eta = 1$, with maximum negativity at $\alpha = -0.2695$. Deviating from this value leads to rapid decay, whereas $-1 \leq \alpha \leq -0.2695$ results in a slowed decay rate. At $\alpha = -0.2695$, entanglement remains stable. Figure 9 shows negativity for $\gamma = 0.2$ at $\eta = 1$, peaking at $\alpha = -0.1206$, where a lower $\alpha$ limit significantly delays entanglement reduction. Even though there is a clear and significant difference in the Hamiltonian and the system employed in this research compared to those used in similar studies, it remains feasible to draw comparisons between the findings of the research cited in [4-11] references and the results obtained in the present study. This means that, despite the variations in the mathematical frameworks applied, a meaningful analysis can still be conducted that relates the outcomes of both investigations, highlighting both congruences and discrepancies that may arise due to the differing Hamiltonian formulations. This research shares an important key result with the studies that have been mentioned previously. Despite the various structural differences that exist among these studies, they all converge on one significant finding. Specifically, they indicate that when the spin chain of the environment is subjected to the process of quantum phase transition, there is a notable increase in the sudden death of entanglement across different systems. This effect is observed in two-qubit systems, as well as in two-qutrit and qubit-qutrit systems, highlighting a consistent trend across these diverse types of quantum states. Holographic negativity is an important way to measure entanglement in quantum systems, helping to understand how information is represented in holographic theories. In systems with two qutrits quantum units that can exist in three states—holographic negativity measures the entanglement between these two subsystems. Current research focuses on exploring how this entanglement connects to the geometric properties of holographic codes, which are mathematical structures used in these theories. The goal is to study the dynamics that result from the interaction between entanglement and geometry. This research includes calculating logarithmic negativity, which is a significant measure of entanglement, particularly relevant for quantum error-correcting codes. By investigating these relationships, researchers aim to deepen our understanding of quantum entanglement and its geometric aspects, contributing to the broader field of quantum information theory.

## References

[1] M. A. Nielsen and I. L. Chuang;   Quantum Computation and Quantum Information; Cambridge University Press, Cambridge (2000).
https://doi.org/10.1017/CBO9780511976667
[2] H. P. Breuer and  F. Petruccione ;" The Theory of open quantum systems”; Oxford University Press, Oxford, New York (2002).
doi: 10.1093/acprof:oso/9780199213900.001.0001
[3] L. Aolita, F. de Melo and L. Davidovich;“Open-system dynamics of entanglement”;Rep. Prog. Phys. 78 (2015) 042001-1-79.

doi: 10.1088/0034-4885/78/4/042001
[4] Z.Sun, X. Wang, C. P. Sun;"Disentanglement in a quantum critical environment"; Phys. Rev. A 75 (2007) 062312.
https://doi.org/10.1103/PhysRevA.75.062312
[5] M.L. Hu ;" Disentanglement dynamics of interacting two qubits and two qutrits in an XY spin-chain environment with the Dzyaloshinsky-Moriya interaction";Phys. Lett. A. 347 (2010) 3520.
https://doi.org/10.1016/j.physleta.2010.06.026
[6] D. Rossini, T. Calarco, V.S.Montangero, R. Fazio;" Decoherence induced by interacting quantum spin baths";Phys. Rev. A 75 (2007) 032333. https://doi.org/10.1103/PhysRevA.75.032333
[7] Y. Yang, W. An-Min;" Correlation dynamics of a qubit—qutrit system in a spin-chain environment with Dzyaloshinsky—Moriya interaction"; Chi Phy B, 23 (2013) 020307.
doi: 10.1088/1674-1056/23/2/020307
[8] C.Y. Lai, J.T. Hung, C.Y. Mou, P.C. Chen ;"Induced decoherence and entanglement by interacting quantum spin baths";Phys. Rev. B 77 (2008) 205419.
https://doi.org/10.1103/PhysRevB.77.205419
[9] J. Jing, Z.G. Lu,;" Dynamics of two qubits in a spin bath with anisotropic XY coupling"; Phys. Rev. B 75 ( 2007) 174425.
https://doi.org/10.1103/PhysRevB.75.174425
[10] Z. G. Yuan, P. Zhang, S. S. Li;" Disentanglement of two qubits coupled to an XY spin chain: Role of quantum phase transition"; Phys. Rev. A 76 (2007) 042118-1-7. https://doi.org/10.1103/PhysRevA.76.042118
[11] L.J. Tian, C.-Y. Zhang, L.-G. Qin ;" SuddenTransition in QuantumDiscord Dynamics: Role of Three-Site Interaction"; Chin. Phys. Lett. 30 (2013) 050303.
doi: 10.1088/0256-307X/30/5/050303
[12] A. Peres;"Separability Criterion for Density Matrices"; Phys. Rev. Lett. 77 (1996) 1413-1415.
https://doi.org/10.1103/PhysRevLett.77.1413
[13] M. Horodecki, P. Horodecki, and R. Horodecki; "Separability of mixed states: necessary and sufficient conditions"; Phys. Lett. A 223 (1996) 1-8.
https://doi.org/10.1016/S0375-9601%2896%2900706-2
[14] G. Vidal and R. F. Werner;"A computable measure of entanglement";Phys. Rev. A 65 (2002) 032314-1-14.
https://doi.org/10.1103/PhysRevA.65.032314
[15] Rangamani, M., Rota, M. Comments on entanglement negativity in holographic field theories. *J. High Energ. Phys.* 2014, 60 (2014).
https://doi.org/10.1007/JHEP10(2014)060
[16] Chaturvedi, P., Malvimat, V. & Sengupta, G. Holographic quantum entanglement negativity. *J. High Energ. Phys.* 2018, 172 (2018). https://doi.org/10.1007/JHEP05(2018)172
[17] Dong, X., Qi, XL. & Walter, M. Holographic entanglement negativity and replica symmetry breaking. *J. High Energ. Phys.* **2021**, 24 (2021). https://doi.org/10.1007/JHEP06(2021)024